\RequirePackage{lineno}
\documentclass[%
 reprint,
twocolumn,
superscriptaddress,
showpacs,preprintnumbers,showkeys,
 amsmath,amssymb,
 aps,
prc,
floatfix,
]{revtex4}

\usepackage{graphicx}% Include figure files
\usepackage{dcolumn}% Align table columns on decimal point
\usepackage{bm}% bold math
\usepackage{amssymb}
\usepackage{float}
\usepackage[caption = false]{subfig}
\usepackage{hyperref}% add hypertext capabiliti
\begin{document}

%\preprint{APS/123-QED}
            
\title{Energy independent scaling of ridge and final state description of high multiplicity p+p collisions at $\sqrt{s}$ = 7 and 13 TeV}
\author{Debojit Sarkar}

\email{debojit03564@gmail.com}
%\author{Subhasis Chattopadhyay}
%\email{sub@vecc.gov.in}
\affiliation{Variable Energy Cyclotron Centre, HBNI, 1/AF-Bidhannagar, Kolkata-700064, India }
\affiliation{Laboratori Nazionali di Frascati, INFN, Frascati, Italy}

%\date{\today}

\begin{abstract}
An energy independent scaling of the near-side ridge yield at a given multiplicity has been observed by the ATLAS and the CMS collaborations in p+p collisions at $\sqrt{s}$ = 7 and 13 TeV. Such a striking feature of the data can be successfully explained by approaches based on initial state momentum space correlation generated due to gluon saturation. In this paper, we try to examine if such a scaling is also an inherent feature of the approaches that employ strong final state interaction in p+p collisions. We find that hydrodynamical modeling of p+p collisions using EPOS 3 shows a violation of such scaling. The current study can, therefore, provide important new insights on the origin of long range azimuthal correlations in high multiplicity p+p collisions at the LHC energies.

%fuels the existing debate on applicability of hydrodynamics in small collision systems.

\end{abstract}

%\pacs{24.30.Cz, 29.40.Mc, 24.60.Dr.}
%\begin{keyword}


%\end{keyword}
%\end{frontmatter}

\maketitle

\section{Introduction} 
The long-range ridge like structure of the azimuthal correlations as observed in the high multiplicity p+p \cite {CMS_pp_Ridge,CMS_pp_v2} and p+A collisions\cite {alice_pPb_double_ridge,pPb_mass_ordering} at the LHC energies shows striking similarity with similar measurements in heavy ion collisions. In heavy ion collisions \cite {HI_review, PbPb_spectra, Au_Au_p_to_Pi, Pb_Pb_P_to_Pi} such long-range correlations are attributed to hydrodynamical response of a strongly interacting system to the initial spatial anisotropy. However, the origin of such correlations in small collision systems are debated to be either driven by hydrodynamic response of initial state geometry \cite{bozek_pPb,epos_ridgein_pp,epos_massordering_flow_pPb,enrg_depen_pp_epos,epos_radialflow_spectra_pPb} or due to intrinsic momentum space correlations among the initially produced patrons \cite{CGC_pPb,PYTHIA_massordering_Tribedy,CGC_ridge_scaling,AmirHRezaeian}. Assuming the applicability of hydrodynamics \cite {Shuryak_radialflow,flow_pp_pghosh,flow_pp_victor,pp_glauber_enterria} one can explain several experimental observations in high multiplicity p+p and p+Pb collisions at LHC energies \cite {bozek_pPb,epos_ridgein_pp, epos_massordering_flow_pPb,epos_radialflow_spectra_pPb}. For example, simulations based on EPOS 3 which includes event by event 3+1 D hydrodynamic evolution can provide  good explanation of the energy dependence of d$N_{ch}$/d$\eta$ in p+p collisions \cite{enrg_depen_pp_epos}, ridge \cite {epos_ridgein_pp,epos_massordering_flow_pPb},  mass ordering of the elliptic flow co-efficients ($v_{2}$) of identified particles \cite {epos_massordering_flow_pPb}, hardening of spectra with multiplicity and the trend of baryon to meson enhancement at the intermediate transverse momentum ($p_{T}$) in p+p and p+Pb collisions at the LHC energies \cite {epos_radialflow_spectra_pPb,alice_pPb_radialflow,p_pi_enhancement_pPb,alice_protontopion_SQM2016,
pp_pPb_meanpt_ALICE}. Also, the hydrodynamical approach implemented inside EPOS 3 can consistently explain the multiplicity dependence of $v_{2}$  in heavy ion collisions as well as in small collision systems \cite{epos_pPb_PbPb_flow,epos_Pb_Pb_ONLY} - providing a unified description of the reaction dynamics for different collision systems at the LHC energies \cite{epos_pPb_PbPb_flow}. Alternative approaches based on purely initial state models and percolation of strings along with quenching of the partons can also explain several qualitative and quantitative features of the ridge like correlations over a wide range of systems and energies.~\cite{CGC_1,CGC_2,CGC_5,CGC_6,CGC_7,CGC_8,CGC_9,CGC_10,
CGC_11,CGC_12,percolation}. Models such as CGC+PYTHIA~\cite{PYTHIA_massordering_Tribedy} has demonstrated that initial state correlations in momentum space generated as a consequence of gluon saturation can survive fragmentation and explain the mass ordering of flow harmonics, mean transverse momentum ($<p_{T}>$) in high multiplicity p+p collisions at the LHC energy.\\ 
By far, no experimental observable has been able to convincingly discriminate the two scenarios. Therefore, the debate, on whether collective behaviors observed in high multiplicity p+p collisions are purely initial state effects driven by gluon saturation \cite{PYTHIA_massordering_Tribedy} or hydrodynamic response to initial geometry \cite{bozek_pPb,epos_ridgein_pp} is still on. In this work we study one particular aspect of the ridge-like correlations that might shed some light into such debate and elucidate the origin of long range correlations in small collision systems.\\ 
The ATLAS and the CMS collaborations have demonstrated an interesting energy independent scaling of the near side ridge like correlations when plotted against the produced charged particle multiplicity in p+p collisions at $\sqrt{s} =$ 7 and 13 TeV \cite{CMS_pp_Ridge_scaling,ATLAS_pp_Ridge_scaling}. In other words, the strength of the near side ridge depends only on the produced multiplicity and not on collision energy. This scaling has been shown to be a natural consequence of the multiparticle production in the regime of gluon saturation \cite{CGC_ridge_scaling}. In a CGC based calculation, in high energy p+p collisions, the saturation scale determines both the multiplicity and the near-side correlated ridge yield and also their energy dependence \cite{CGC_ridge_scaling}. Therefore, fixing the multiplicity also fixes the ridge yield and one obtains an universal scaling curve between ridge yield and multiplicity at different energies as shown in Ref \cite{CGC_ridge_scaling}. By far, such scaling has not been explored using an approach that employ strong final state interaction. In this paper, we therefore explore such scaling using the hydrodynamical framework as implemented in EPOS 3. Any indication of a possible violation in the energy independent scaling of ridge yield in EPOS 3 will provide important insight about the origin of ridge like structure in high multiplicity p+p collisions.\\ 
%In particular, such observation will also constrain the relation between produced multiplicity and the initial spatial anisotropy.
 The goal of this work is to qualitatively investigate the experimentally observed energy independent scaling of ridge yield in hydrodynamical framework rather than quantitatively explain the data. The paper is organised as follows. In the next section, we give a brief description of the EPOS-3 event generator. The construction of the correlation function and the method of extraction of the ridge yield are discussed in section 3. The discussions on the results and a summary are provided in the final section.

\section{ EPOS 3 Model}

The EPOS 3 model includes a 3+1 D hydrodynamical simulation with flux tube initial conditions \cite {epos_ridgein_pp,epos_radialflow_spectra_pPb}  \cite {epos_model_descrip,epos_core_corona_sep,trigdilution_EPOS_pPb,ridge_EPOS_pPb_jetmedium_int}. For a detailed description of EPOS 3 we refer the reader to Refs \cite {epos_ridgein_pp,epos_radialflow_spectra_pPb,epos_massordering_flow_pPb}  \cite {parton_gribov_Regge_Th}. The initial state dynamics used in this model is based on ``Parton based Gribov Regge Theory" \cite {parton_gribov_Regge_Th}. After multiple initial state scatterings, a partonic system is produced that consists of longitudinal colour flux tubes or strings (known as ``pomerons") \cite {epos_ridgein_pp} of small transverse size (radius $\approx$ 0.2 fm) which are extended upto many units of space-time rapidity $\eta_{s}$. The multiplicity of an event is proportional to the number of initially produced pomerons (flux tubes / strings). They carry transverse momentum of the hard scattered partons in the transverse direction. Depending on the energy of the string segments and local string density - the high density area form a ``core" and the low density area form a``corona" region\cite {epos_core_corona_sep}. In a typical low-multiplicity p+p collisions, one expects the formation of a corona dominated system in which the hadrons are produced via string fragmentation (Schwinger mechanism) \cite {epos_model_descrip,epos_massordering_flow_pPb}.
But, in high multiplicity p+p collisions, due to large number of initial parton-parton scatterings, the string density becomes high enough that the strings cannot decay independently \cite{epos_ridgein_pp}. Instead, one switches to a final state description in which the energy density and the flow velocity are estimated from the four-momenta of the infinitesimal string segments and used as initial condition for the final hydrodynamic evolution \cite{epos_ridgein_pp}. In EPOS 3, only the ``core" region (dominant in high multiplicity events) undergoes hydrodynamic evolution \cite {epos_massordering_flow_pPb} and finally hadrons are produced from such region via  Cooper-Frye prescription.\\ 
The ridge like structure in the high multiplicity p+p collisions in EPOS 3 is generated in the follwing way. After multiple parton-parton scatterings in the initial state, the transverse positions (spatial) of the flux tubes are generated randomly and overlapping of these flux tubes generate the high density core region. The fluctutations in the transverse positions of the overlapping flux tubes may lead to an eccentric (dominantly elliptic) shape of the core region and the hydrodynamic response to such eccentricity eventually leads to a cos(2$\Delta\phi$) modulation in the azimuthal correlation function \cite {epos_massordering_flow_pPb}. The longitudinal invariance of the initial state of EPOS 3 is responsible for the long-range ridge like structure of such azimuthal correlation. EPOS 3 also generates di-jet correlations through the corona region. The strings in the corona hadronize by Schwinger's mechanism and constitutes the jet part of the produced system. 
This model has successfully explained several qualitative and quantitative features of the ridge like  structures observed in high multiplicity p+p \cite {epos_ridgein_pp} and p+Pb \cite {epos_massordering_flow_pPb} collisions at the LHC energies. In this paper, we will use EPOS 3 to study the two particle correlation and investigate the experimentally observed energy independent scaling of ridge yields \cite {CMS_pp_Ridge_scaling,ATLAS_pp_Ridge_scaling,CGC_ridge_scaling} in p+p collisions at $\sqrt{s}$ = 7 and 13 TeV.

\begin{figure}[htb!]
\begin{center}

a)

\includegraphics[scale=0.46]{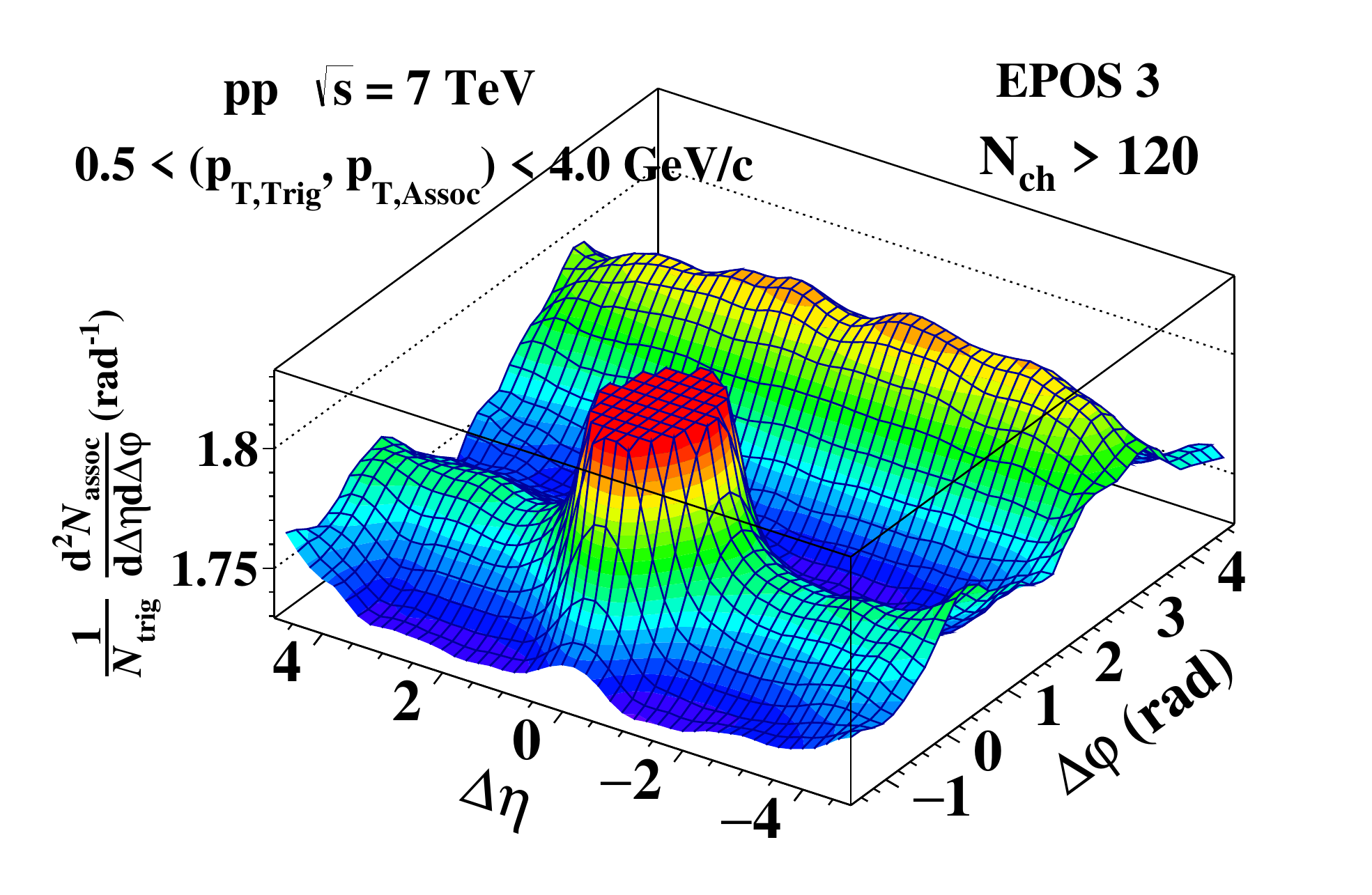}

b)

\includegraphics[scale=0.46]{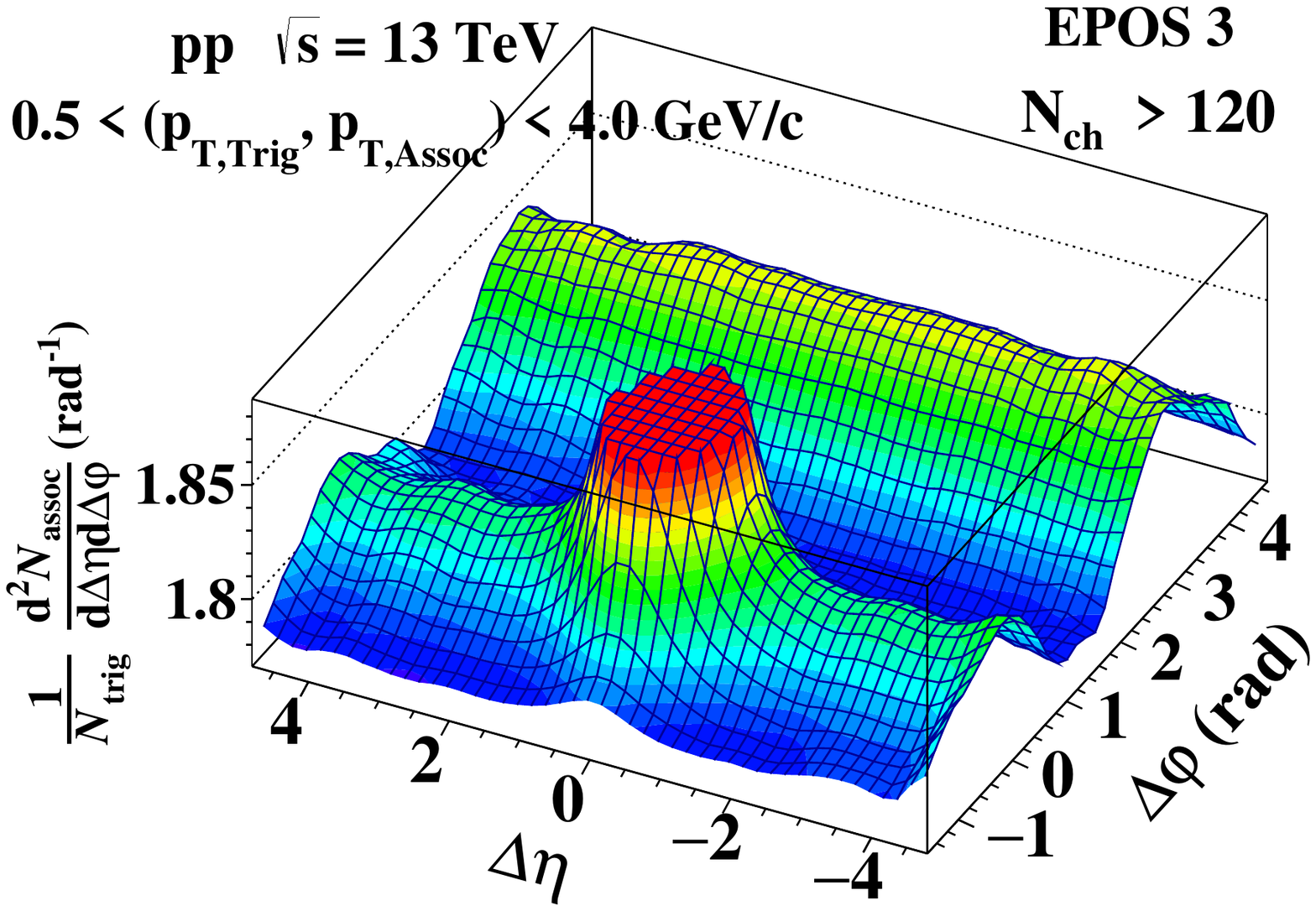}
\caption{[Color online] Two particle $\Delta\eta$-$\Delta\phi$ correlation function as obtained from EPOS-3 
in p+p collisions for $N_{ch} > 120$ at a) $\sqrt{s}$ = 7 TeV and b) $\sqrt{s}$ = 13 TeV.}
\end{center}
\label{corr-p-pi}
\end{figure}

\section{Analysis Method}
In the present analysis, the correlation function is obtained 
among two sets of particles classified as {\it trigger} and {\it associated}. The p${_T}$ range of trigger and 
associated particles are 0.5  $<p{_T}<$ 4.0 GeV/$\it{c}$ and the correlation function has been constructed with a $p_{T}$ ordering ($p_{T}^{Assoc} < p_{T}^{Trigger}$). The pseudorapidity of the particles are 
restricted within -2.4$<\eta<$2.4 and the reason for such choice is motivated by the CMS acceptance \cite {CMS_pp_Ridge_scaling}. A two dimensional (2D) correlation function is obtained as a function of relative
azimuthal angle $\Delta\phi$ = $\bf\phi_{trigger}$ -$\bf\phi_{associated}$ and relative pseudorapidity $\Delta\eta$
= $\bf\eta_{trigger}$ -$\bf\eta_{associated}$. The same event correlation function is defined as 
$\frac{dN_{same}}{N_{trigger}d\Delta\eta d\Delta\phi}$, where $N_{same}$ is the number of particles associated to 
triggers particles ($N_{trigger}$) that are taken from the same event. To correct for pair acceptance the same event correlation function is divided by mixed event correlation function  $\alpha$$\frac{dN_{mixed}}{d\Delta\eta d\Delta\phi}$. The mixed event correlation function is constructed by correlating the trigger particles in one event with the associated particles from other events within the same multiplicity class. The factor $\alpha$ is used to normalize  the mixed event to make it unity for pairs where both particles go into approximately the same direction ($|\Delta\eta|$ $\approx$ 0, $|\Delta\phi|$ $\approx$ 0) \cite{trigdilution_EPOS_pPb,ridge_EPOS_pPb_jetmedium_int}.\\
\begin{figure}[htb!]
\begin{center}
\includegraphics[height=6.0 cm, width=8.4 cm]{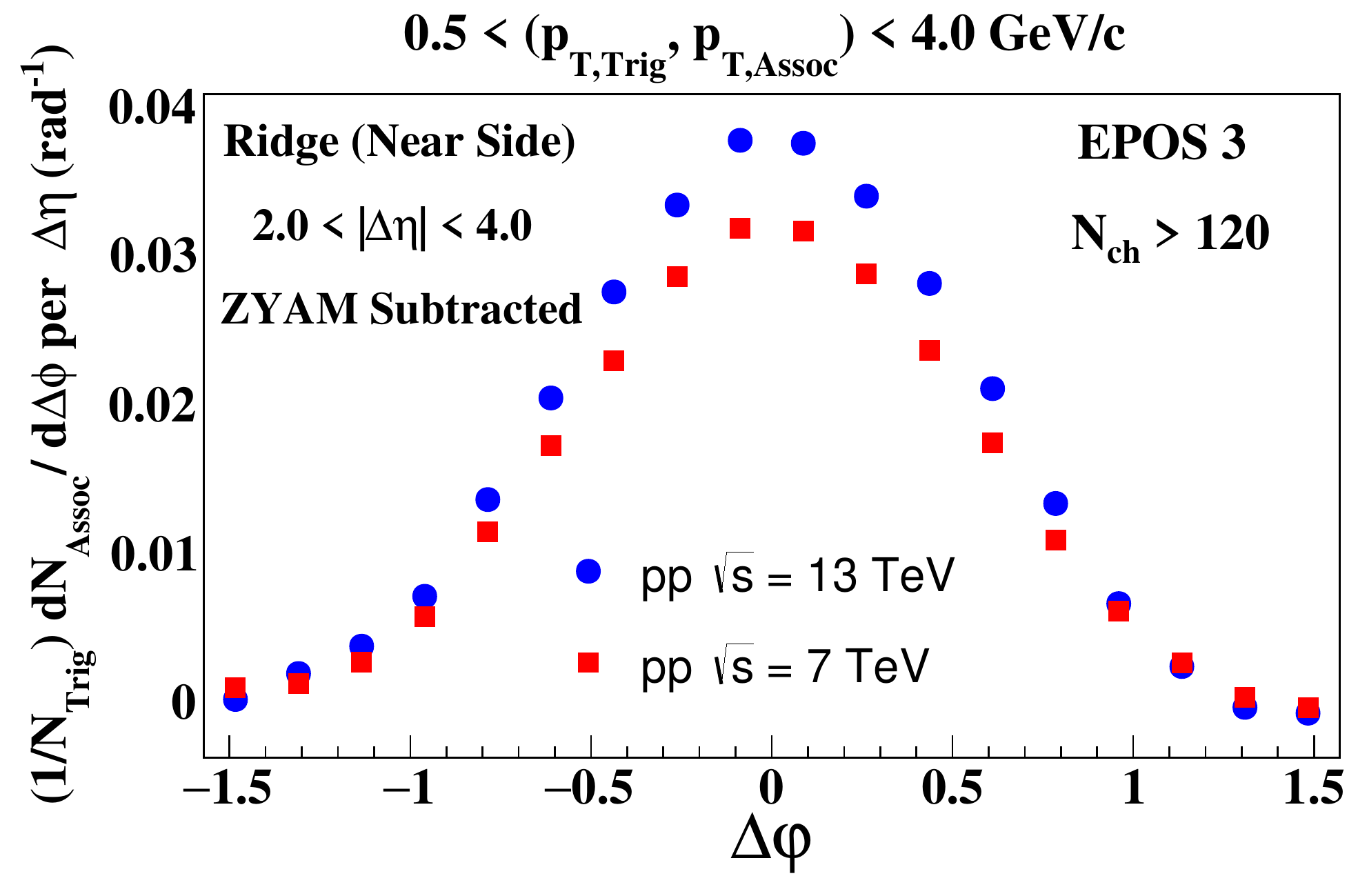}
\caption{[Color online] The comparison of the ZYAM subtracted $\Delta\varphi$ projection of the long range (2.0 $ <|\Delta\eta| <$ 4.0) near side ($|\Delta\phi|$  $< \pi/2$) correlation function in p+p collisions at $\sqrt{s}$ = 7 TeV and $\sqrt{s}$ = 13 TeV. }
\end{center}
\label{Yield_ratio_Pr_Pi}
\end{figure}

In this analysis, the correlation functions are constructed in 6 multiplicity classes
defined by the number of particles ($N_{ch}$) produced within $|\eta| < $ 2.4 and $p_{T} >$ 0.4 GeV/c  {ref}. The 2D correlation functions with ($N_{ch} >$ 120) in pp collisions at 7 and 13 TeV are shown in Fig 1. The particles from jet fragmentation are expected to be confined in a small angular region ($|\Delta\eta|$ $\approx$ 0, $|\Delta\phi|$ $\approx$ 0)- so the ridge is estimated from the $\Delta\phi$ projection over the range (2.0 $ <|\Delta\eta| <$ 4.0). The pedestal is determined from the $\Delta\phi$ projection with the zero yield at minimum (ZYAM) assumption \cite {ridge_EPOS_pPb_jetmedium_int,ZYAM} and subtracted from the correlation function. The ZYAM subtracted $\Delta\phi$ projection of the near side ($|\Delta\phi|$  $< \pi/2$) ridge estimated from the 2.0 $ <|\Delta\eta| <$ 4.0 is shown in Fig 2. The near side ridge yield is estimated by integrating the
$\Delta\phi$ projection in the range  $|\Delta\phi|$  $< \pi/2$ \cite {ridge_EPOS_pPb_jetmedium_int}.

\begin{figure}[htb!]
\begin{center}
\includegraphics[scale=0.43]{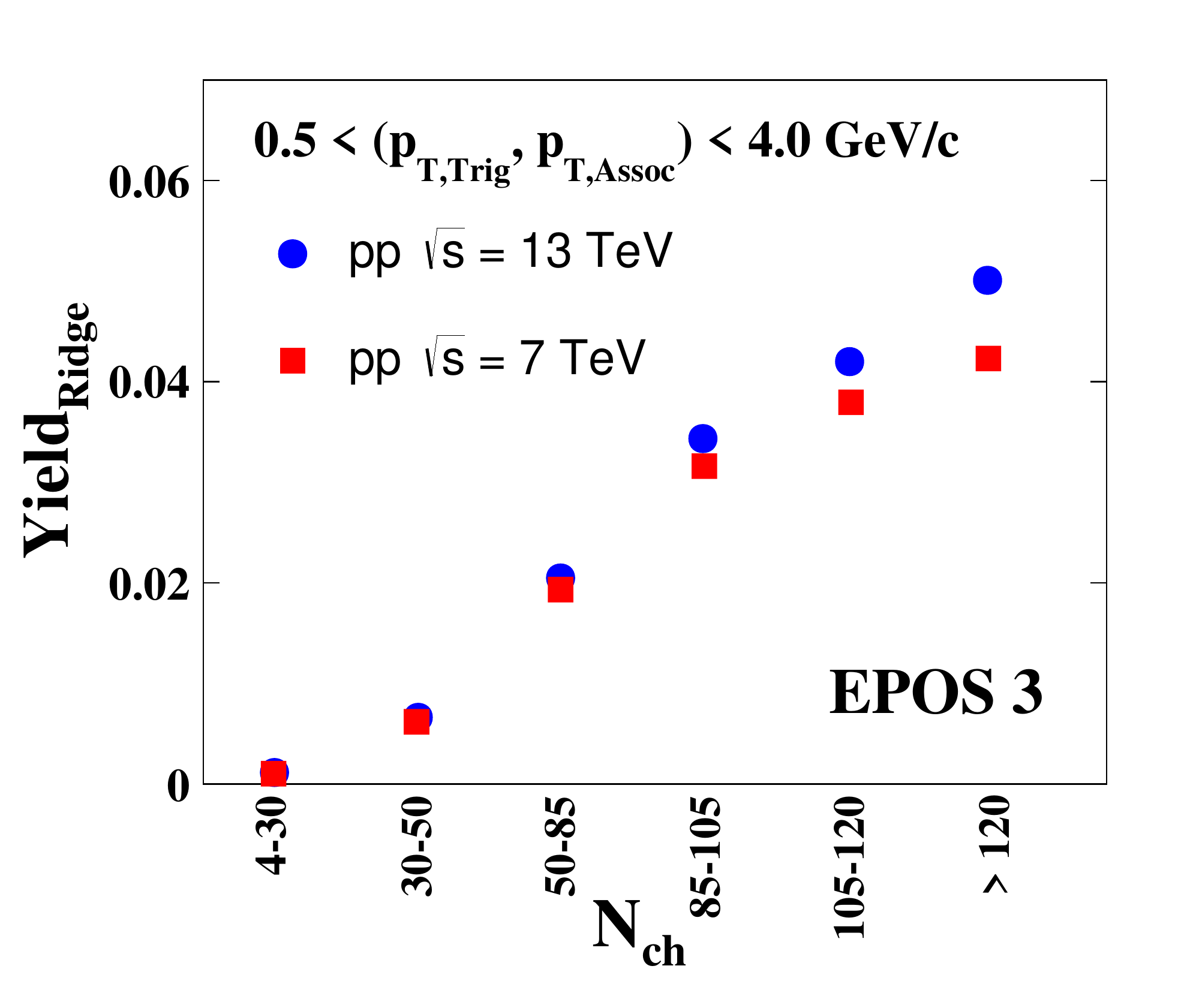}
\caption{[Color online]  Multiplicity dependence of the near-side ($|\Delta\phi|$  $< \pi/2$) ridge (2.0 $ <|\Delta\eta| <$ 4.0) yield (ZYAM subtracted) as estimated from EPOS-3 in p+p collisions at $\sqrt{s}$ = 7 TeV and $\sqrt{s}$ = 13 TeV.}
\end{center}
\label{Yield_ratio_Pr_Pi}
\end{figure}

\section{Results and Discussion} 
Aforementioned, in current work we restrict ourselves to only qualitative comparisons of the di-hadron correlations with the experimental data \cite{CMS_pp_Ridge_scaling,ATLAS_pp_Ridge_scaling}. In Fig.1 the 2D $\Delta\eta$-$\Delta\phi$ correlation functions estimated from EPOS 3 in the highest multiplicity class ($N_{ch} >$ 120) for p+p collisions at $\sqrt{s} =$ 7 and 13 TeV are shown. The long range ridge structure along with the jet peak is observed in both the cases similar to the experimental data. For the purpose of this work,
\begin{figure}[htb!]
\begin{center}
\includegraphics[scale=0.43]{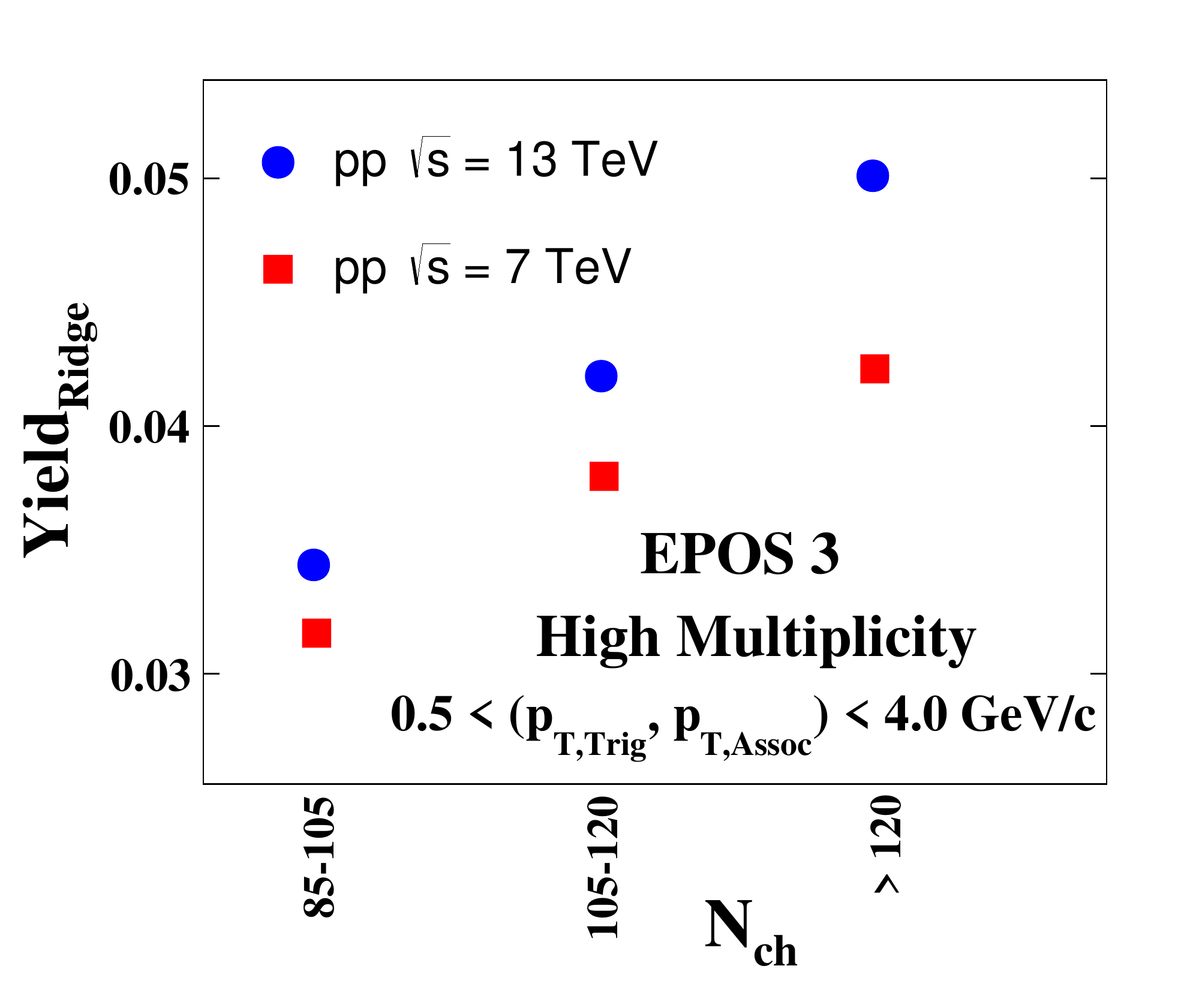}
\caption{[Color online] Multiplicity dependence of the near-side ($|\Delta\phi|$  $< \pi/2$) ridge (2.0 $ <|\Delta\eta| <$ 4.0) yield (ZYAM subtracted) as estimated from EPOS-3 in high multiplicity ($N_{ch}> 85$) p+p collisions at $\sqrt{s}$ = 7 TeV and $\sqrt{s}$ = 13 TeV.}
\end{center}
\label{Yield_ratio_Pr_Pi}
\end{figure}
we concentrate only on the long range (2.0 $ <|\Delta\eta| <$ 4.0) of the near side ($|\Delta\phi|$  $< \pi/2$) ridge component of the correlation function and therefore make a $\Delta\phi$ projection over the range 2.0 $ <|\Delta\eta| <$ 4.0 as shown in Fig. 2. The strength of the ridge in p+p collisions at 13 TeV is found to be stronger compared to that of p+p collisions at 7 TeV. So, Fig. 2 indicates that at a fixed multiplicity, the anisotropy in the azimuthal distribution of particles is higher at higher collision energies.\\
In order to better demonstrate this effect, we integrate the ZYAM subtracted long range (2.0 $ <|\Delta\eta| <$ 4.0) correlation function over the range ($|\Delta\phi|$  $< \pi/2$) to extraxct the near side ridge yield for different multiplicity bins. Fig. 3 shows the multiplicity evolution of the near side ridge yield in EPOS-3 for p+p collisions at both energies for 0.5 $ <(p_{T,Trig}, p_{T,Assoc}) <$ 4.0 GeV/c. At lower multiplicities ($N_{ch} <$ 85), the ridge yield is found to be approximately energy independent. However, at higher multiplicities, a strong energy dependence of the ridge yield is observed as highlighted in Fig. 4. The ridge yield in p+p collisions at 13 TeV is found to be consistently higher compared to the p+p collisions at 7 TeV for $N_{ch} >$ 85. In other words, we don't see any energy independent scaling of ridge yield with multiplicity in EPOS 3 as observed by the ATLAS and the CMS collaborations ~\cite{CMS_pp_Ridge_scaling,ATLAS_pp_Ridge_scaling}. Such scaling has been observed to hold in the LHC data even for multiplicity classes that correspond to ten times the mean multiplicity \cite{CMS_pp_Ridge_scaling,ATLAS_pp_Ridge_scaling}. A possible reason for this scaling violation in higher multiplicity classes ($N_{ch} >$ 85) of EPOS 3 can be understood as follows. At higher multiplicity classes, the core which originates due to the overlapping of the flux tubes constitutes the dominant part of the system \cite{ridge_EPOS_pPb_jetmedium_int, ortiz_PYT_EPOS}. It is therefore expected that the final state effects possibly leads to the observed scaling violation. 
As discussed previously, in EPOS 3, the anisotropy in the azimuthal distribution of the final state particles originates from the hydrodynamic response to the initial anisotropy (eccentricity) of the transverse energy density profile of the core region \cite{epos_ridgein_pp}.  Such eccentricity is generated from the statistical fluctuations in the transverse positions of the initially produced flux tubes, an approach similar to the one discussed in \cite{wiedemann_fluxtube}. The produced multiplicity of the final state particles, on the other hand, does not depend on the details of such fluctuations. Therefore, nothing prevents the initial eccentricity and final multiplicity to have different energy dependence. One, therefore, does not expect the strength of ridge, driven by initial eccentricity, to be independent of energy at a fixed multiplicity.\\ 
Now, the scaling violation appears to be prominent only in the core dominated higher multiplicity classes. In EPOS 3, the multiplicity (or the energy) per individual flux tube increases with $\sqrt{s}$.
Due to higher initial energy density at higher collision energies, a smaller overlapping region (``core") can contribute to the same multiplicity originating from a larger overlap region in lower energy collisions. The initial transverse energy density is expected to be more anisotropic in the case of less overlap between the fluxtubes compared to the case where overlap is more \cite {bozek_fluxtube}. As a result, at a fixed multiplicity, higher eccentricity, i.e. a stronger ridge is expected in higher collision energies compared to the lower collision energy case.

In summary, the observation of long range ridge like correlations in p+p collisions has been a topic of great interest in recent times. Several features of the ridge-like correlations have been both qualitatively and quantitatively explained by approaches that are based on both initial state effects due to gluon saturation and final state effects like hydrodynamic evolution in the presence initial spatial anisotropy. In this work, we investigate a very specific experimental observation which indicates that the strength of the near side ridge yield follows an energy independent scaling when plotted against the multiplicity of produced particles in p+p collisions at LHC energies  \cite{CMS_pp_Ridge_scaling,ATLAS_pp_Ridge_scaling}. Such energy independent scaling of ridge yield at the LHC has been explained from an approach based on gluon saturation in which the multiparticle production is determined from a single saturation scale \cite{CGC_ridge_scaling,CMS_pp_Ridge_scaling,ATLAS_pp_Ridge_scaling}. In this work we demonstrate that the approaches based on strong final state effects do not show such scaling. We argue that such violation of energy independent scaling constraints the interpretations of long range correlation in high multiplicity p+p collisions based on final state effects at the LHC energies. More specifically, to observe an energy independent scaling of the ridge with multiplicity in final state approach, one needs to incorporate additional constraints such as energy independent scaling of initial eccentricity with produced multiplicity.

\iffalse
In short, the current study can provide important insights about the final state based description of long range azimuthal correlations in high multiplicity p+p collisions at LHC energies. A violation in the energy independent scalling of the ridge yield at fixed multiplicity has been observed in high multiplicity classes of EPOS 3 (final state approach). Therefore, an energy independent scaling of ridge with multiplicity, as observed in the experiment \cite{CGC_ridge_scaling,CMS_pp_Ridge_scaling,ATLAS_pp_Ridge_scaling}, naturally constrains the energy dependence of the initial eccentricity and the produced multiplicity. Further investigation on the energy dependence of such scaling at LHC energies using different final state approaches will be essential to constrain those models aiming to explain the observed collective like behaviors in small collision systems.
\fi

\section*{Acknowledgements} 
I acknowledge fruitful discussions and suggestions from Prithwish Tribedy on this manuscript. I also thank Subhasis Chattopadhyay for his help and support throughout this work. I thank Klaus Werner for allowing me to use EPOS 3 for this study. Thanks to the VECC grid computing team for their constant effort to keep the facility running and helping in EPOS data generation and data analysis.

%In particular, such observation will also constrain the relation between produced multiplicity and the initial %spatial anisotropy.

\iffalse
the eccentricity increases with the decrese in the number of flux tubes.  At a fixed multiplicity, as the number of flux tubes are less at higher collision energy, a stronger ridge (due to larger eccentricity) is expected compared to the lower collision energy case.

Therefore, less number of flux tubes can contribute to the same multiplicity at higher $\sqrt{s}$ as compared to that of lower collision energies. 
\fi
%So, different collision energies can produce similar multiplicity from different initial eccentricities.

\end{document}